%% file: nfp.tex
\documentclass{article} 
\usepackage[accepted]{icml2017}
\usepackage{times}

\usepackage[paperVersion]{marcpaper}

\icmltitlerunning{Differentiable Functional Program Interpreters}

\newcommand{\terpret}{\textsf{Terp\reflectbox{r}\reflectbox{e}\reflectbox{t}}}

\begin{document}
\twocolumn[
\icmltitle{Differentiable Functional Program Interpreters}


\newcommand{\fix}{\marginpar{FIX}}
\newcommand{\new}{\marginpar{NEW}}



\begin{icmlauthorlist}
\icmlauthor{John K.\ Feser}{MIT}
\icmlauthor{Marc Brockschmidt}{MSR}
\icmlauthor{Alexander L. Gaunt}{MSR}
\icmlauthor{Daniel Tarlow}{Google}
\end{icmlauthorlist}

\icmlaffiliation{MIT}{Massachusetts Institute of Technology, Cambridge, US}
\icmlaffiliation{MSR}{Microsoft Research, Cambridge, UK}
\icmlaffiliation{Google}{Google Brain, Montr\'{e}al, Canada (work done while at Microsoft)}

\icmlcorrespondingauthor{John K.\ Feser}{feser@csail.mit.edu}

\icmlkeywords{program induction}

\vskip 0.3in
]

\printAffiliationsAndNotice{}

\begin{abstract}
  Programming by Example (PBE) is the task of inducing computer programs from
  input-output examples. It can be seen as a type of machine learning where the
  hypothesis space is the set of legal programs in some programming language.
  Recent work on \emph{differentiable interpreters} relaxes the discrete space
  of programs into a continuous space so that search over programs can be performed
  using gradient-based optimization.
  While conceptually powerful, so far differentiable
  interpreter-based program synthesis has only been capable of solving very
  simple problems.
  In this work, we study modeling choices that arise when constructing a
  differentiable programming language and their impact on the success of
  synthesis.
  The main motivation for the modeling choices comes
  from functional programming: we study the effect of
  memory allocation schemes, immutable data, type systems, and built-in
  control-flow structures.
  Empirically we show that incorporating functional programming ideas
  into differentiable programming languages allows us to learn much more complex
  programs than is possible with existing differentiable languages.
\end{abstract}

\input{introduction}
\input{relatedwork}

\input{diffinterp}

\input{model}
\input{experiments}
\input{discussion}



\newpage
\bibliography{references}
\bibliographystyle{icml2017}

\appendix
\input{appendix}

\end{document}

%% file: introduction.tex
\section{Introduction}
\label{sect:introduction}

A key decision in supervised machine learning is to choose a hypothesis
space, which is the space of possible mappings from inputs to outputs.
Common choices for hypothesis spaces are neural networks, decision trees, and nearest
neighbor models.
The premise underlying this work is that programs,\footnote{In this work, by ``program'' we mean a program that is represented as source code in a human-readable programming language.} or programs in combination with other models like neural networks,
are a useful hypothesis class for machine learning.
Programs are interesting models because (1) they can be very expressive---they
define complex modern technical infrastructure like operating systems
and the internet---and (2) they come with a strong inductive bias.
Programming languages are designed to make common programming patterns compact
and easy to express, and with appropriate priors this might bias a learning method to favor hypotheses that humans consider to be natural programs.
The ultimate goal of this line of work is to build models that
generalize strongly from a small amount of data but are expressive enough
to fit large, complex data sets. Such models might be particularly applicable to
sequential and procedural data.

However, recent results indicate that programs are a difficult hypothesis class
to work with. There is significant literature on discrete search-based techniques
for program induction (e.g., \cite{Gulwani11,recursive-synthesis,Feser15,frankle2016}), and a small amount of recent work
on gradient-based program induction \cite{Bunel16,Riedel16,Gaunt16}. \citep{Gaunt16} has
shown for low-level assembly-like differentiable programming languages,
discrete search performs better than gradient-based search, and further that the
kinds of problems that can be solved by differentiable interpreters are limited
to simple problems like accessing an element of an array.

One may then wonder why it is worth continuing to study differentiable programming
languages. We believe there are three main reasons. First, differentiable programming
languages allow program-like models to be composed with neural network-like
models, which opens up many possibilities. However, the bottleneck
in scaling up is that differentiable interpreters cannot currently solve complex
problems.
Second, differentiable interpreters appear fundamentally different from discrete search, and they may have benefits.
For example, differentiable interpreters naturally operate in the stochastic regime with small minibatches of data processed at each step, whereas discrete search methods typically operate in a batch regime.\footnote{Or in counterexample-guided inductive synthesis (CEGIS) setting, where data instances are added to a monotonically growing active set.} They may also
 give more natural ways of handling noisy data. The related final point is that
differentiable interpreters are a very new development, and they have
not been studied nearly as extensively as discrete search techniques.
We believe it important to study them in different contexts, to better understand where their strengths and weaknesses lie.
In this work, we focus on differentiable interpreters for higher level programming languages than have previously been studied.

Specifically, in this work we show that ideas from modern high-level functional programming languages
can be used to improve differentiable interpreters.
We show how to adapt several key ideas and discuss why they
lead to more powerful differentiable interpreters.
In total, we develop a functional programming language
operating on integers and lists and a corresponding differentiable interpreter.
In our empirical evaluation, we show the effects on learning performance of
our four modeling recommendations, namely
 automatic memory management,
 the use of combinators and if-then-else constructs to structure program control flow,
 immutability of data, and
 an application of a simple type system.
Our experiments show that each of these features crucially improves program
learning over existing baselines.


%% file: relatedwork.tex
\section{Related Work}
\label{sect:relatedwork}

\paragraph{Inductive Program Synthesis}
There has been significant recent interest in synthesizing functional programs from input-output examples in the programming languages community.
Synthesis systems generally operate by searching for a program which is correct on the examples, using types or custom deduction rules to eliminate parts of the search space.
Among the notable systems: \textsc{Myth}~\citep{osera2015,frankle2016} synthesizes recursive functional programs from examples using types to guide the search for a correct program,
$\lambda^2$~\citep{Feser15} synthesizes data structure manipulating programs structured using combinators using types and deduction rules in its search,
\textsc{Escher}~\citep{recursive-synthesis} synthesizes recursive programs using search and a specialized method for learning conditional expressions,
and FlashFill~\citep{Gulwani11} structures programs as compositions of functions and uses custom deduction rules to prune candidate programs.
Our decision to learn functional programs was strongly inspired by this previous work.
In particular, the use of combinators to structure control flow was drawn from~\cite{Feser15}.
The key difference in our work is the use of differentiable interpreters and gradient-based optimization instead of the discrete search employed in the above works.

\eat{
A related line of research is the extension of neural network architectures with
components that correspond to hardware
primitives~\citep{Giles89,Graves14,Weston14,Joulin15,Grefenstette15,Kurach15,Kaiser15,Reed15,Andrychowicz16,Zaremba16,Graves16},
enabling them to learn program-like behavior.
However, these models are usually tightly coupled to the idea of a
differentiable interpretation of computer hardware, as names such as
 Neural Turing Machine~\citep{Graves14},
 Neural Random-Access Machine~\citep{Kurach15}, and
 Neural GPU~\citep{Kaiser15}
indicate.
We observe that while such architectures form the basis modern computing, they
are usually not the models that are used to program computers.
Instead, decades of programming languages research have lead to ever higher
programming languages that aim to make programming simpler and less error-prone.
Indeed, as recent comparisons show~\citep{Gaunt16}, program synthesis methods
from the programming languages community that actively exploit such
constructs, e.g. by leveraging known semantics of loops, are currently achieving
considerably better results than comparable neural architectures.
Still, neural IPS techniques are clearly at an advantage when extending the
problem setting from simple integer input/output examples to more complex cases,
such as IPS problems with perceptual data~\citep{Gaunt16b}, imprecise examples,
or leveraging additional cues such as a natural language description of the
desired program.
}
\vspace{-1ex}
\paragraph{Neural Networks Learning Algorithms}
A number of recent models aim to learn algorithms from input/output data.
Many of these augment standard (recurrent) neural network architectures with
differentiable memory and simple computation components (e.g.
\cite{Graves14,Kurach15,Joulin15,Neelakantan15,Graves16}).
The main commonality between differentiable interpreters and these works is
the smoothing technique that is used to convert discrete operations into
a continuous parameterization.
The main difference is that these models use a neural network controller
to decide which operation to perform next, whereas differentiable interpreter-based
models store the decision of what operations to perform in the source code itself,
in conjunction with some kind of \emph{instruction pointer} that denotes where in
the source code the current execution is.
\citet{Zaremba16} induce algorithms using a reinforcement learning setup,
which avoids the need for the smoothing operations. Like the other above works, it uses a neural network
controller to decide the order in which to perform operations.
\citet{Reed15} learn algorithms from strong supervision specifying which operation to perform at each step. \citet{li2017neuralprogram} weakens the supervision requirement somewhat
but still requires supervision in the form of sequences of basic actions and
some strong supervision.

None of these works focus on producing source code, and thus we
do not expect them to achieve the same strong generalization properties that come
from using a source code representation.
In experiments, we will show that the source code-based models can generalize from 5
examples, whereas a model using a neural network-based controller fails to do so.

Finally, the most related work is that on differentiable interpreters.
\citet{Bunel16} and \citet{Riedel16} have used program models
similar to assembly (resp. Forth) source code to initialize solutions, and
either optimize or complete them.
\citet{Gaunt16} develops a framework that allows comparing differentiable
interpreters to several alternative synthesis systems, focusing on low-level
programming models including Turing machines, Boolean circuits, and an assembly-like
language.
Our work differs from these in that we focus on the question of how to design
a programming model to improve the performance of differentiable interpreter-based
synthesis.


%% file: diffinterp.tex
\newcommand{\maxTime}{\ensuremath{\mathit{T}}\xspace}
\newcommand{\timeStep}{\ensuremath{\mathit{t}}\xspace}

\newcommand{\maxInt}{\ensuremath{\mathit{M}}\xspace}

\newcommand{\maxLen}{\ensuremath{\mathit{P}}\xspace}
\newcommand{\progPos}{\ensuremath{\mathit{p}}\xspace}

\newcommand{\maxInstr}{\ensuremath{\mathit{I}}\xspace}
\newcommand{\instr}{\ensuremath{\mathit{i}}\xspace}
\newcommand{\instrPtr}{\ensuremath{\mathit{p}}\xspace}
\newcommand{\branch}{\ensuremath{\mathit{b}}\xspace}

\newcommand{\maxReg}{\ensuremath{\mathit{R}}\xspace}
\newcommand{\outReg}{\ensuremath{\mathit{o}}\xspace}
\newcommand{\fstRegArg}{\ensuremath{\mathit{a}_1}\xspace}
\newcommand{\sndRegArg}{\ensuremath{\mathit{a}_2}\xspace}
\newcommand{\condRegArg}{\ensuremath{\mathit{c}\xspace}}
\newcommand{\regVal}{\ensuremath{\mathit{r}}\xspace}

\newcommand{\maxStack}{\ensuremath{\mathit{H}}\xspace}
\newcommand{\stackVal}{\ensuremath{\mathit{h}}\xspace}

\newcommand{\regType}{\ensuremath{\leftidx{^t}{\mathit{r}}\xspace}}

\newcommand{\nextState}{\ensuremath{\eta}\xspace}
\newcommand{\state}{\ensuremath{\mathit{s}}\xspace}

\newcommand{\probEq}[2]{\ensuremath{\llbracket #1 = #2 \rrbracket}}

\section{Differentiable Interpreters}
\label{sect:repr}

We begin by reviewing differentiable interpreters and
defining a basic program and data representation to introduce the general concepts.

Programs operate on states consisting of an instruction pointer indicating
the next instruction to execute, a number of registers holding inputs and
intermediate results of executed instructions, and a heap containing memory
allocated by the program.
We focus on list-manipulating programs, so we use a heap represented as an array
of (data, pointer) value pairs, where pointers are indices of this array, or the
special 0 value.
To represent a linked list, each cell points to the next cell in the list, except for the last cell, which points to 0.
The elements of the list are stored in the data part of each heap location.

\paragraph{Program Representation. }
All of our models share a basic instruction set, namely
 the constructor \code{cons} which stores a (data, pointer) value
  pair on the heap and returns a pointer to the newly created heap cell,
 the heap accessors (\code{head} \& \code{tail}) which return the data
  (resp. pointer) element of a heap cell,
 integer addition, increment and decrement (\code{add}, \code{inc},
  \code{dec}),
 integer equality and greater-than comparison (\code{eq} \& \code{gt}),
 Boolean conjunction and disjunction (\code{and} \& \code{or}),
 common constants (\code{zero} \& \code{one}), and
 finally a \code{noop} instruction.
 These all have the expected semantics,
 although we will discuss the behavior of \code{cons} in detail later.

We pick
 a maximal integer value \maxInt, 
 a number of instructions \maxInstr,
 and a number of registers \maxReg.
In this introductory setting, the size of the heap memory \maxStack has to be equal to the
maximal integer value \maxInt, but we will relax this later.

We limit the length of programs to some value \maxLen, and can then encode
programs as a sequence of tuples
 $(\outReg^{(\progPos)}, \instr^{(\progPos)}, \fstRegArg^{(\progPos)}, \sndRegArg^{(\progPos)})$,
where 
 $\instr^{(\progPos)} \in [1, \maxInstr]$ identifies the \progPos-th
instruction and 
 $\outReg^{(\progPos)}, \fstRegArg^{(\progPos)}, \sndRegArg^{(\progPos)} \in [1, \maxReg]$
its output and argument registers respectively.

\paragraph{Interpreter. }
An interpreter takes a program description and executes it.
In our setting we limit the number of execution timesteps \maxTime and keep a
program state
 $\state^{(\timeStep)} =
   (\instrPtr^{(\timeStep)},
    \regVal_1^{(\timeStep)} \ldots \regVal_\maxReg^{(\timeStep)},
    \stackVal_1^{(\timeStep)} \ldots \stackVal_\maxStack^{(\timeStep)})$
for each timestep \timeStep, where $\instrPtr^{(\timeStep)} \in [1, \maxLen]$ is
an instruction pointer indicating which instruction to execute next,
$\regVal^{(\timeStep)}_\ast$ are the values of registers, and
$\stackVal^{(\timeStep)}_\ast$ are the values of the (pointer, data) cells in the heap.
The interpreter works by iteratively updating the program state by executing
the next instruction, which is determined by the instruction pointer.
For example, executing \code{(\outReg, \code{add}, \fstRegArg,
  \sndRegArg)} on a state at timestep $\timeStep$ yields the following registers
at the next timestep:
\begin{align*}
 \regVal_u^{(\timeStep + 1)}
  & = 
     \begin{cases}
       \regVal_{\fstRegArg}^{(\timeStep)} + \regVal_{\sndRegArg}^{(\timeStep)} \quad \mathit{mod}~\maxInt
          & \text{if } u = \outReg\\
       \regVal_u^{(\timeStep)}
          & \text{otherwise.}
     \end{cases}
  & \forall u \in [1, \maxReg]
\end{align*}

\paragraph{Differentiable Interpreter. }
To make an interpreter differentiable, we follow earlier work (e.g. \citep{Graves14,Kurach15,Bunel16,Riedel16,Gaunt16})
and replace all discrete values with probability distributions over their values and
lift all operations to operate on probability distributions instead of discrete values
by averaging over all the possibilities with weights given by the probability distributions.
For example, if
 $\nextState(\state^{(\timeStep)}, (\outReg, \instr, \fstRegArg, \sndRegArg))$
computes the state obtained by executing the instruction
 $(\outReg, \instr, \fstRegArg, \sndRegArg)$,
we can compute the next state $\state^{(\timeStep+1)}$ as follows, where
\probEq{x}{n} denotes the probability that a variable $x$ encoding a discrete
probability distribution assigns to the value $n$.
\begin{align}
 \state^{(\timeStep+1)} =\hspace{-5ex}
  \sum_{\substack{
        \instrPtr \in [1, \maxLen],
        \instr \in [1, \maxInstr],\\
        \outReg, \fstRegArg, \sndRegArg \in [1, \maxReg]}}
         \begin{array}{l}
                    \probEq{\instrPtr^{(\timeStep)}\!}{\!\instrPtr}
              \cdot \probEq{\outReg^{(\instrPtr)}\!\!}{\!\outReg}
              \cdot \probEq{\instr^{(\instrPtr)}\!\!}{\!\instr}\\\quad
              \cdot \probEq{\fstRegArg^{(\instrPtr)}\!\!}{\!\fstRegArg}
              \cdot \probEq{\sndRegArg^{(\instrPtr)}\!\!}{\!\sndRegArg}\\\quad
              \cdot \nextState(\state^{(\timeStep)}\hspace{-1ex}, (\outReg, \instr, \fstRegArg, \sndRegArg))
         \end{array}
 \label{eq:evalStep}
\end{align}
We can then differentiate with respect to the parameters of these probability distributions.
To do program synthesis, we randomly initialize the representation of the program (a collection of probability distributions over discrete values) and then use gradient ascent to find distributions over program
variables that (locally) maximize the log probability of observed (discrete) output values given observed (discrete) input values.
In a bit more detail, our aim is to learn the program parameters
 $(\outReg^{(\progPos)}, \instr^{(\progPos)}, \fstRegArg^{(\progPos)}, \sndRegArg^{(\progPos)})$
such that program ``evaluation'' according to \rEq{eq:evalStep} starting on a
state $s^{(0)}$ initialized to an example input yields the target output in
$s^{(\maxTime)}$.
For scalar outputs such as a sum of values, our objective is simply to minimize
the cross-entropy between the distribution in the \emph{output} register
$r_{\maxReg}^{(\maxTime)}$ and a point distribution with all probability mass on
the correct output value.
In practice, we developed our models in \terpret. See
\citet{Gaunt16} for more details.

%% file: model.tex
\section{Differentiable Functional Program Interpreters}
\label{sect:model}

In the following, we will discuss how and why to add functional programming
features to differentiable interpreters.
We start with the simple assembly-like language from the previous section
and progress to a differentiable version of a simple functional programming
language.
We begin by developing an observation model for list-structured data, and then
we make four modeling recommendations inspired by functional programming.
Empirically, we will demonstrate the benefit of these recommendations
in \rSC{sect:experiments}.

\subsection{Observation Model for List Data}

We first discuss a new observation model for list-structured data.
Handling list outputs is more complex than scalar values, as there are
many ways for a target list to be represented in memory.
Intuitively, we will traverse the heap from the returned heap address
until reaching the end of a linked list, recording the list elements as we go,
and then we will observe the sequence of elements that was recorded.
To formalize this intuition, let
  $\leftidx{^d}\stackVal^{(\maxTime)}_{k}$ 
  (resp. $\leftidx{^p}\stackVal^{(\maxTime)}_{k}$)
denote the data (resp. pointer) information in the heap cell at address $k$ at
the final state of the evaluation.
We then compute the traversal sequences of list element 
 values $v_1, \ldots, v_{\maxStack}$ and
 addresses $a_1, \ldots, a_{\maxStack}$
as follows.
\begin{align*}
    a_i &= \begin{cases}
            r_{\maxReg}^{(\maxTime)} & \text{if } i = 1 \\
            \sum_{a \in [1, \maxStack]} \probEq{a_{i-1}}{a} \cdot \leftidx{^p}\stackVal^{(\maxTime)}_{a} & \text{otherwise}
          \end{cases}\\
    v_i &= \sum_{a \in [1, \maxStack]} \probEq{a_i}{a} \cdot \leftidx{^d}\stackVal^{(\maxTime)}_{a}
\end{align*}
The probability that the computed output list is equal to an expected output
list $[\bar{v}_1, \ldots, \bar{v}_k]$ is then $\probEq{a_{k+1}}{0} \cdot
\sum_{i=1}^k \probEq{v_i}{\bar{v}_i}$.

\subsection{Structured Control Flow}

\begin{wrapfigure}[5]{r}{4.3cm}
\vspace{-4ex}
\hspace*{-4ex}
\begin{tikzpicture}[terpretmodel]
 \node[constant] (l1)  at (0,0)                          {\code{1:}};
 \node[outpar]  (out1) at ($(l1.east) + (0.2,0)$)        {};
 \node[constant] (eq1) at ($(out1.east) + (0.1,0)$)      {$\leftarrow$};
 \node[instr] (instr1) at ($(eq1.east) + (0.1,0)$)       {};
 \node[arg1] (arg11)   at ($(instr1.east) + (0.2,0)$)    {};
 \node[arg2] (arg21)   at ($(arg11.east) + (0.2,0)$)     {};
 \node[branch] (br1)   at ($(arg21.east) + (0.2,0)$)     {};
 \node[constant] (l2)  at ($(l1.south west) + (0,-0.4)$) {\code{2:}};
 \node[outpar]  (out2) at ($(l2.east) + (0.2,0)$)        {};
 \node[constant] (eq2) at ($(out2.east) + (0.1,0)$)      {$\leftarrow$};
 \node[instr] (instr2) at ($(eq2.east) + (0.1,0)$)       {};
 \node[arg1] (arg12)   at ($(instr2.east) + (0.2,0)$)    {};
 \node[arg2] (arg22)   at ($(arg12.east) + (0.2,0)$)     {};
 \node[branch] (br2)   at ($(arg22.east) + (0.2,0)$)     {};
 \node[constant]       at ($(arg12.south west) + (0,-.4)$) {$\ldots$};
\end{tikzpicture}
\vspace{-5ex}
\caption{\label{fig:ASMOutline}}
\end{wrapfigure}
Our baseline program model corresponds closely to an assembly language as
used in earlier work~\citep{Bunel16}, resulting in a program model as shown
in \rF{fig:ASMOutline}, where boxes correspond to learnable parameters.
We extend our instruction set with jump-if-zero (\code{jz}),
jump-if-not-zero (\code{jnz}) and \code{return} instructions.
Our assembly program representation also includes a ``branch''
parameter $\branch$ specifying the new value of the instruction pointer for a
successful conditional branch.
To learn programs in this language, the model must learn how to create the
control flow that it needs using these simple conditional jumps.
Note that the instruction pointer suffers from the same problems as
the stack pointer above, i.e., uncertainty about its value blurs together the
effects of many possible program executions.

\begin{wrapfigure}[10]{r}{5.3cm}
\vspace{-3ex}
\hspace*{-3.5ex}
\begin{tikzpicture}[terpretmodel]
 \node[constant] (lp1)  at (0,0)                             {$\mathit{pre}_1$:};
 \node[outpar]  (outp1) at ($(lp1.east) + (0.2,0)$)          {};
 \node[constant] (eqp1) at ($(outp1.east) + (0.1,0)$)        {$\leftarrow$};
 \node[instr] (instrp1) at ($(eqp1.east) + (0.1,0)$)         {};
 \node[arg1] (arg1p1)   at ($(instrp1.east) + (0.2,0)$)      {};
 \node[arg2] (arg2p1)   at ($(arg1p1.east) + (0.2,0)$)       {};
 \node[cond] (condp1)   at ($(arg2p1.east) + (0.2,0)$)       {};
 \node[constant] (lp2)  at ($(lp1.south west) + (0, -.25)$)  {$\mathit{pre}_2$:};
 \node[constant]        at ($(arg1p1.south west) + (0,-.4)$) {$\ldots$};
 \node[constant] (ll)   at ($(lp1.west) + (0,-1)$)           {\code{foreach ele in}};
 \node[loopVar]  (lv)   at ($(ll.east) + (0.1,0)$)           {};
 \node[constant] (ll1)  at ($(ll.south west) + (0.5,-.35)$)  {$\mathit{loop}_1$:};
 \node[outpar]  (outl1) at ($(ll1.east) + (0.2,0)$)          {};
 \node[constant] (eql1) at ($(outl1.east) + (0.1,0)$)        {$\leftarrow$};
 \node[instr] (instrl1) at ($(eql1.east) + (0.1,0)$)         {};
 \node[arg1] (arg1l1)   at ($(instrl1.east) + (0.2,0)$)      {};
 \node[arg2] (arg2l1)   at ($(arg1l1.east) + (0.2,0)$)       {};
 \node[cond] (condl1)   at ($(arg2l1.east) + (0.2,0)$)       {};
 \node[constant] (ll2)  at ($(ll1.south west) + (0, -.25)$)  {$\mathit{loop}_2$:};
 \node[constant]        at ($(arg1l1.south west) + (0,-.4)$) {$\ldots$};
 \node[constant] (ls1)  at ($(ll.south west) + (0, -1.4)$)   {$\mathit{suf}_1$:};
 \node[outpar]  (outs1) at ($(ls1.east) + (0.2,0)$)          {};
 \node[constant] (eqs1) at ($(outs1.east) + (0.1,0)$)        {$\leftarrow$};
 \node[instr] (instrs1) at ($(eqs1.east) + (0.1,0)$)         {};
 \node[arg1] (arg1s1)   at ($(instrs1.east) + (0.2,0)$)      {};
 \node[arg2] (arg2s1)   at ($(arg1s1.east) + (0.2,0)$)       {};
 \node[cond] (conds1)   at ($(arg2s1.east) + (0.2,0)$)       {};
 \node[constant] (ls2)  at ($(ls1.south west) + (0, -.25)$)  {$\mathit{suf}_2$:};
 \node[constant]        at ($(arg1s1.south west) + (0,-.4)$) {$\ldots$};
\end{tikzpicture}
\vspace{-5.5ex}
\caption{\label{fig:LoopOutline}}
\end{wrapfigure}
A key difference between hardware-level assembly languages and higher-level
programming languages is that higher-level languages structure control flow using loops, conditional
statements, and procedures, as raw \texttt{goto}s are famously considered harmful~\citep{dijkstra1968}.
Functional languages go a step further and leverage higher-order functions to
abstract over common control flow patterns such as iteration over a recursive data
structure.
In an imperative language,
such specialized control flow is often repeated and mixed with other
code.
In the differentiable interpreter setting, structured control flow gives an additional
benefit, which is that it reduces uncertainty in the instruction pointer.

To introduce structured control flow, we replace raw jumps with an \code{if-then-else} instruction and
an explicit \code{foreach} loop that is suited for processing lists.
We restrict our model to a prefix of instructions, a loop which iterates over a list,
and a suffix of instructions.
The parameters for instructions in the loop can access an additional register
that contains the value of the current list element.
An outline of such a program is shown in \rF{fig:LoopOutline}.
This removes uncertainty about the value of the instruction pointer, as each
time step corresponds to exactly one ``line'' in the program template.
To implement this behavior in practice, we unroll the loop for a fixed number of
iterations derived from the bound on the size of the input, which ensures that every
input list can be processed.

For the \code{if-then-else} instruction, we extend the instruction
representation with a ``condition'' parameter $\condRegArg \in [1, \maxReg]$ and
let the evaluation of \code{if-then-else} yield its first argument when
the register $\condRegArg$ is non-zero and the second argument otherwise.
An overview of the structure of such programs is displayed above.


\begin{wrapfigure}[11]{l}{4.4cm}
\vspace{-3.5ex}
\hspace*{-1.5ex}
\begin{tikzpicture}[terpretmodel]
  \node[constant]   (s01)  at (0,0)                                {\code{foldli}:};
  \node[constant]   (s11)  at ($(s01.south west) + (0,-0.4)$)                                {\code{acc}};
  \node[constant]   (s12)  at ($(s11.east) + (0.1,0)$)             {$\leftarrow$};
  \node[register]   (s13)  at ($(s12.east) + (0.1,0)$)             {};
  \node[constant]   (s21)  at ($(s11.south west) + (0,-0.2)$)      {\code{idx}};
  \node[constant]   (s22)  at ($(s21.east) + (0.1,0)$)             {$\leftarrow$};
  \node[constant]   (s23)  at ($(s22.east) + (0.1,0)$)             {\code{0}};
  \node[constant]   (s31)  at ($(s21.south west) + (0,-0.2)$)      {\code{foreach ele in}};
  \node[loopVar]    (s32)  at ($(s31.east) + (0.1,0)$)             {};
  \node[outpar]     (s71)  at ($(s31.south west) + (0.6,-0.4)$)    {};
  \node[constant]   (s72)  at ($(s71.east) + (0.1,0)$)             {$\leftarrow$};
  \node[instr]      (s73)  at ($(s72.east) + (0.1,0)$)             {};
  \node[arg1]       (s74)  at ($(s73.east) + (0.2,0)$)             {};
  \node[arg2]       (s75)  at ($(s74.east) + (0.2,0)$)             {};
  \node[cond]       (s76)  at ($(s75.east) + (0.2,0)$)             {};
  \node[constant]   (s81) at ($(s71.south west) + (0,-0.3)$)       {$\ldots$};
  \node[constant]   (s91)  at ($(s81.south west) + (-0.15,-0.3)$)   {\code{acc}};
  \node[constant]   (s92)  at ($(s91.east) + (0.1,0)$)             {$\leftarrow$};
  \node[register]   (s93)  at ($(s92.east) + (0.1,0)$)             {};
  \node[constant]   (s101)  at ($(s91.south west) + (0,-0.2)$)     {\code{idx}};
  \node[constant]   (s102)  at ($(s101.east) + (0.1,0)$)           {$\leftarrow$};
  \node[constant]   (s103)  at ($(s102.east) + (0.1,0)$)           {\code{idx + 1}};
  \node[outpar]     (s111)  at ($(s101.south west) + (-0.3,-0.3)$) {};
  \node[constant]   (s112)  at ($(s111.east) + (0.1,0)$)           {$\leftarrow$};
  \node[constant]   (s113)  at ($(s112.east) + (0.1,0)$)           {\code{acc}};
\end{tikzpicture}
\vspace{-4.5ex}
\caption{\label{fig:FoldliOutline}}
\end{wrapfigure}
We note that while fixing the iteration over the list elements is already helpful,
learning most list-processing programs requires the model to repeatedly infer the concepts of
creating a new list, aggregating results and keeping track of the current list
index.
In functional programming languages, such regular patterns are encapsulated in
combinators.
Thus, in a second model, we replace the simple \code{foreach} loop with three
combinators:
 \code{mapi} creates a new list by applying a function to each element
 of the input list,
 \code{zipWithi} creates a new list by iterating over two lists in parallel and
 applying a function to both elements, and
 \code{foldli} computes a value by iterating over all list elements and applying
 a function to the current list element and the value computed so far.
A program model using the \code{foldli} combinator is shown in
\rF{fig:FoldliOutline}.
The \code{i} suffix indicates that these combinators additionally provide the
index of the current list element (the precise semantics of the combinators are
presented in \rSC{sect:appendix:combinators}).

\textbf{Recommendation (L):} Instead of raw jumps, use loop and if-then-else
templates.

\subsection{Memory Management}
Most modern programming languages eschew manual memory management and pointer
manipulation where possible.
Instead, creation of heap objects automatically generates an appropriate
pointer to fresh memory.
Similarly, built-in constructs allow access to fields of objects, instead of
requiring pointer arithmetic.
Both of these choices move program complexity into the fixed implementation of
a programming language, making it easier to write correct programs.

As the programs we want to learn need to construct new lists, we explored two
memory allocation mechanism that provides fresh cells.
First, we attempted to use a \emph{stack pointer} $sp$ which always points to
the next free memory cell, and fixing a maximum stack size \maxStack.
Whenever a memory cell is allocated (i.e., a \code{cons} instruction is
executed), the stack pointer is incremented.

There are two problems with this allocation mechanism.
1)~We must maintain a copy of the heap and stack pointer for each timestep \timeStep.
For large values of \maxTime or \maxStack, this significantly increases the size of the model.
2)~Uncertainty about whether an instruction is \code{cons} translates
into uncertainty about the precise value of the stack pointer, as each call to
\code{cons} changes $sp$.
This uncertainty causes cells holding results from different instructions in the
stack to blur together, despite the fact that cells are immutable once created.
As an example, consider the execution of two instructions, where the first is 
 \code{cons 1 0} with probability $0.5$ and \code{noop} otherwise,
and the second is
 \code{cons 2 0} with probability $0.5$ and \code{noop} otherwise.
After executing starting with $sp = 1$, the value of
$sp$ will be blurred across three values $1$, $2$ and $3$ with probabilities
$0.25$, $0.5$ and $0.25$.
Similarly, the value of the first heap cell will be $0$ (the default) with
probability $0.25$, $1$ with probability $0.5$ and $2$ with probability $0.25$.
This blurring effect becomes stronger with longer programs, and we found that it
substantially impacted learning.

Both of these problems can be solved by transitioning to a fully immutable
representation of the heap.
In this variant, we allocate and initialize one heap cell per timestep, i.e., we
set $\maxStack = \maxTime$.
If the current instruction is a \code{cons}, the appropriate values are filled
in, otherwise both data and pointer value are set to a default value (in our
case, $0$).
This eliminates the issue of blurring between outputs of different
instructions.
The values of a heap cell may still be uncertain as they inherit uncertainty
about the executed instructions and the values of arguments, but depend only on
the operations at one timestep.
Because there is no uncertainty about whether to fill in a heap cell, we keep a single copy of the heap and use the current timestep \timeStep as the stack pointer.
While this modification requires a larger domain to store pointers, we found not
copying the stack significantly reduces memory usage during training of our
models.

\textbf{Recommendation (F):} Use fixed heap memory allocation
deterministically controlled by the model.

\subsection{Immutable Data}
In functional programming, functions are expected to not have side effects,
and all data is immutable.
This helps programmers reason about their code, as it eliminates the possibility
that a variable might be left uninitialized or accessed in an inconsistent state.
Moreover, no data is ever ``lost'' by being overwritten or mutated.

In training initial models, we observed that many random initializations of the
program parameters would overwrite input data or important intermediate
results.
In models with combinators that provide a way to accumulate result
values, we can sidestep this issue by making registers immutable.
To do so, we create one register per timestep, and fix the output
of each instruction to the register for its timestep.
Parameters for arguments then range over all registers initialized in prior
timesteps, with an exception for the closures executed by a combinator.
Here, each instruction only gets access to the inputs to the closure, values computed in the
prefix, and registers initialized by preceding instructions in the same loop
iteration.
As in the heap allocation case, we can avoid keeping a copy of all
registers for every timestep, and instead share these values over all steps.
A somewhat unintuitive consequence is that this strategy reduces memory
footprint of the model.

\textbf{Recommendation (I):} Use immutable registers by deterministically
choosing where to store outputs.

\subsection{Types}

In programming languages, expressive type systems are used to protect programmers from writing programs that will fail.
Practically, a type checker is able to rule out many syntactically correct
programs that are certain to fail at runtime, and thus restricts the space of valid
programs.
When training initial models, we found that for many initializations, training would
fall into local minima corresponding to ill-typed programs, e.g., where
references to the heap would be used in integer additions.
We expect the learned program to be well-typed,
so we introduce a simple type system.
We explored two approaches to adding a type system.

Our first attempt extended the objective with a penalty for type errors.
In our programs, we use three simple types of data---integers, pointers and
booleans---as well as a special type, $\bot$, which represents type errors.
We extended the program state to contain an additional element
$\regType$ for each register, encoding its type.
Each instruction then not only computes a value that is assigned to the target
register, but also a type for the target register.
Most significantly, if one of the arguments has an unsuitable type (e.g., an
integer in place of a pointer), the resulting type is $\bot$.
We then extended our objective function to add a penalty for values with type
$\bot$.
Unfortunately, this changed objective function had neither a positive nor
negative effect on our experiments, so it seems that optimizing for
the correct type is redundant when we are already optimizing for the
correct return value.

In our second attempt, rather than penalizing ill-typed programs, we prevent
programs from accessing ill-typed data by construction.
We augment our register representation by adding an integer, pointer, and
Boolean slot to each register, so each register can hold a separate value of
each type.
Instructions which read from registers now read from the slot corresponding to
the type of the argument.
When writing to a register, we write to the slot corresponding to the
instruction's return type, and set the other slots to a default value 0.
This prevents any ill-typed sequence of instructions, i.e., it is now impossible
to, for example, increment a pointer value or to fill the pointer part of a heap cell with a
non-pointer value.
Furthermore, this modification allows us to set the heap size \maxStack to a value
different from the maximal integer \maxInt because it allows pointers and integers to have different maximum values.

\textbf{Recommendation (T):} Use different storage for data of different
types.


%% file: experiments.tex
\section{Experiments}
\label{sect:experiments}

We have empirically evaluated our modeling recommendations on a selection of program
induction tasks of increasing complexity, ranging from straight-line programs to
problems with loops and conditional expressions.
All of our models are implemented in \terpret~\citep{Gaunt16} and we learn using
\terpret{}'s \textsc{TensorFlow}~\citep{Tensorflow} backend.

For all tasks, three groups of five input/output example pairs were sampled as
training data and another 25 input/output pairs as test data.
For each group of five examples, training was started from 100 random
initializations of the model parameters.
After training for 3500 epochs (tests with longer training runs showed no
significant changes in the outcomes), the learned programs were tested by
discretizing all parameters and comparing program outputs on test inputs
with the expected values.
We perform 300 runs per model and task, and report only the ratio of successful runs.
A run is successful if the discretized program returns the correct
result on all five training and 25 test examples.\footnote{We inspected samples
  of the obtained programs as well and verified that they were indeed correct
  solutions. See \rSC{sect:appendix:exampleResults} for some of the
    learned programs.}
The ratio of runs converging to zero loss on the training examples was within
1\% of the number of successful runs, i.e., very few found solutions failed to generalize.

We performed a cursory exploration of hyperparameter choices,
sampling 100 hyperparameter settings (choosing optimization algorithm,
learning rate, gradient noise, (decay of) entropy bonus, and gradient
clipping) and tested their effect on two simple tasks.
We ran the remaining experiments with the best configuration obtained
by this process: the RMSProp optimization algorithm, a learning
rate of $0.1$, clipped gradients at $1$, and no gradient noise.

We consider the ratio of successful runs as earlier work has identified this as
a significant problem.
For example, \cite{Neelakantan15b} reports that even after a (task-specific)
``large grid search'' of hyperparameters, the Neural Random Access Machine
converged only in 5\%, 7\% and 22\% of random restarts on the considered tasks.
Similar observations were made in \cite{Kaiser15,Bunel16,Gaunt16} for related
program learning models.

\newcommand{\cmbName}{{\sc C+T+I}\xspace}
\newcommand{\cmbuntName}{{\sc C+I}\xspace}
\newcommand{\cmbmutName}{{\sc C+T}\xspace}
\newcommand{\cmbuntmutName}{{\sc C}\xspace}
\newcommand{\asmName}{{\sc A}\xspace}
\newcommand{\asmfxaName}{{\sc A+F}\xspace}
\newcommand{\asmloopName}{{\sc A+L}\xspace}
\newcommand{\lambdaSName}{\ensuremath{\lambda^2}\xspace}

In our experiments we evaluate the effect of the choices discussed in
\rSC{sect:model}, comparing seven model variants in total.
We call our initial assembly model \asmName and its variation with a fixed
memory allocation scheme \asmfxaName.
All other models use the fixed memory allocation scheme.
The extension of the assembly model with a built-in \code{foreach} loop is
called \asmloopName.
The model including predefined combinators is called \cmbuntmutName, where
\cmbuntName (resp. \cmbmutName) are its extensions with immutable registers
(resp. typed registers).
Finally, \cmbName combines all of these, making it a simple end-to-end
differentiable functional programming language.

As baselines, we consider \lambdaSName~\citep{Feser15}, a strong
program synthesis baseline from programming languages research, and
an implementation of the Neural Random Access Machine (NRAM)~\cite{Kurach15}.
We chose \lambdaSName because of its built-in support for list-processing
programs.
As \lambdaSName is deterministic, we only report a success ratio of either $1$
(if a program matching all input-output examples was generated) or $0$ (if no
so such program was generated) for all experiments.
We ran \lambdaSName with a timeout of 600 seconds.
We give a detailed description of our NRAM model and experimental results
separately in \rSC{sect:experiments:nram}.

\subsection{Straight-line programs}
\label{sect:experiments:straightline}

\begin{figure*}[t]
\begin{tikzpicture}
\definecolor{cmbColor}{rgb}{0,0,1}
\definecolor{cmbuntColor}{rgb}{0,1,0}
\definecolor{cmbmutColor}{rgb}{1,0,0}
\definecolor{cmbuntmutColor}{rgb}{0,1,1}
\definecolor{asmColor}{rgb}{0.75,0.5,0}
\definecolor{asmfxaColor}{rgb}{0.5,0.75,0}
\definecolor{asmloopColor}{rgb}{0.75,0,0.5}
\definecolor{lambdaSColor}{rgb}{0,0.75,0.5}

\pgfplotsset{
  cmb/.style={color=cmbColor,mark=square*},
  cmbmut/.style={color=cmbmutColor,mark=square},
  cmbunt/.style={color=cmbuntColor,mark=*},
  cmbuntmut/.style={color=cmbuntmutColor,mark=o},
  asm/.style={color=asmColor,mark=triangle*},
  asmfxa/.style={color=asmfxaColor,mark=diamond},
  asmloop/.style={color=asmloopColor,mark=diamond*},
  lambdaS/.style={color=lambdaSColor,mark=pentagon*},
  every axis plot/.append style={line width=1pt},
}

\begin{groupplot}[
                  group style={
                               group size=2 by 1,
                               ylabels at=edge left,
                               yticklabels at=edge left,
                              },
                  tiny,
                  height=4.75cm,
                  width=0.525*\linewidth,
                 ]
  \nextgroupplot[
                 title={\code{dupK}},
                 ylabel={Success ratio},
                 xmin=1, xmax=9,
                 ymin=0, ymax=1,
                 axis on top,
                 ytick={0,0.1,0.2,0.3,0.4,0.5,0.6,0.7,0.8,0.9,1},
                 yticklabels={0.0,,0.2,,0.4,,0.6,,0.8,,1.0},
                 xtick={1,2,3,4,5,6,7,8,9},
                 xticklabels={1,2,3,4,5,6,7,8,9},
                 enlargelimits=0.025,
                ]
  \addplot [cmb] table {dupK_cmb.data};
  \addplot [cmbunt] table {dupK_cmbunt.data};
  \addplot [cmbmut] table {dupK_cmbmut.data};
  \addplot [cmbuntmut] table {dupK_cmbuntmut.data};
  \addplot [asm] table {dupK_asm.data};
  \addplot [asmfxa] table {dupK_asmfxa.data};
  \addplot [asmloop] table {dupK_asmloop.data};
  \addplot [lambdaS] table {dupK_lambda2.data};

  \nextgroupplot[
                 title={\code{getK}},
                 xmin=1, xmax=9,
                 ymin=0, ymax=1,
                 axis on top,
                 ytick={0,0.1,0.2,0.3,0.4,0.5,0.6,0.7,0.8,0.9,1},
                 yticklabels={0.0,,0.2,,0.4,,0.6,,0.8,,1.0},
                 xtick={1,2,3,4,5,6,7,8,9},
                 xticklabels={1,2,3,4,5,6,7,8,9},
                 enlargelimits=0.025,
                 legend entries={{\cmbName},{\cmbuntName},{\cmbmutName},{\cmbuntmutName},{\asmName},{\asmfxaName},{\asmloopName},{\lambdaSName}},
                 legend cell align={left},
                ]
  \addplot [cmb] table {getK_cmb.data};
  \addplot [cmbunt] table {getK_cmbunt.data};
  \addplot [cmbmut] table {getK_cmbmut.data};
  \addplot [cmbuntmut] table {getK_cmbuntmut.data};
  \addplot [asm] table {getK_asm.data};
  \addplot [asmfxa] table {getK_asmfxa.data};
  \addplot [asmloop] table {getK_asmloop.data};
  \addplot [lambdaS] table {getK_lambda2.data};
\end{groupplot}
\end{tikzpicture}
\vspace{-3ex}
\caption{Success ratio of our models on straight-line programs of increasing length}
\label{fig:straightline-results}
\vspace{-2ex}
\end{figure*}
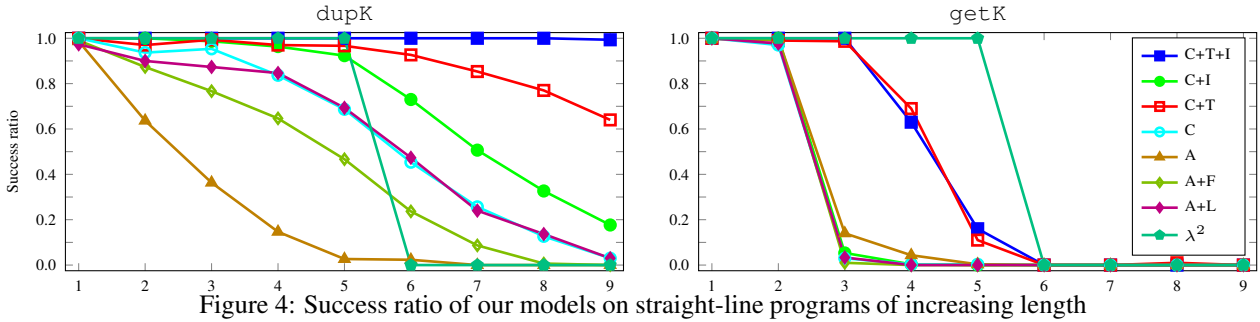

In our first experiment, we consider two families of simple problems---solvable with
straight-line programs---to study the interaction of our modeling choices with
program length.
Our first benchmark task is to duplicate a scalar input a fixed number $k$
times to create a list of length $k$.
Our second benchmark is to retrieve the $k$-th element of a list, again fixing $k$
beforehand (we will consider a generalization of this task where $k$ is a
program input later).
We set the hyperparameters for all models to allow 11 statements, i.e., for \asmName and \asmfxaName we have set
the program length to 11, and for the \asmloopName and \cmbuntmutName\!* models
we have set the prefix and loop length to 0 and the suffix length to 11.
For models where the number of registers does not depend on the number of timesteps,
we use 3 registers, with one initialized to the input.
This allows for $\sim 10^{39}$ programs in the \asmName, \asmfxaName,
\cmbuntName, and \cmbName models and for $\sim 10^{28}$ programs in the
remaining models.
%
These parameters were chosen to be slightly larger than required by the
largest program to be learned.
For all of our experiments, the maximal integer \maxInt was set to $20$ for
models where possible (i.e., for \asmName, \cmbName, \cmbmutName), and to \maxStack
(derived from \maxTime, coming to $22$) for the others.\footnote{We also
  experimented with varying the value of \maxInt. Choices over 20 showed no
  significant differences to smaller values.}

We evaluated all of our models following the regime discussed above and present
the results in \rF{fig:straightline-results} for $k$ values from $1$ to $9$.
The difference between \asmName and \asmfxaName on the \code{dupK} task
illustrates the significance of \textbf{Recommendation (F)} to fix the memory
allocation scheme.
Following \textbf{Recommendation (T)} to separate values of different types
improves the results on both tasks, as the differences between \cmbName (resp.
\cmbmutName) and \cmbuntName (resp. \cmbuntmutName) illustrate.

\subsection{Simple loop programs}
\label{sect:experiments:simple}
In our second experiment, we compare our models on three simple
list algorithms: computing the length of a list, reversing a list and
summing a list.
Model parameters have been set to allow 6 statements for the \asmName and
\asmfxaName models, and empty prefixes, empty suffixes, and 2 instructions in the
loop for the other models. 
For models where the number of registers does not depend on the number of timesteps,
we use 4 registers, with one initialized to the input.

\begin{table*}[t]
  \vspace{-.5ex}
  \caption{Success ratio for experiments on simple loop-requiring tasks.}
  \label{tab:simple-loop-results}
  \vspace{-2ex}
  \begin{centering}
    {\small
     \input{loopy_simple_tbl.tex}

    }
  \end{centering}
  \vspace{-1ex}
\end{table*}

The results of our evaluation are displayed in \rTab{tab:simple-loop-results},
starkly illustrating \textbf{Recommendation (L)} to use predefined loop
structures.
We speculate that learning explicit jump targets is extremely challenging
because changes to the parameters controlling jump target instructions have
outsized effects on all computed (intermediate and output) values.
On the other hand, models that could choose between different list iteration
primitives were able to find programs for all tasks.
We again note the effect of \textbf{Recommendation (T)} to separate values of
different types on the success ratios for the \code{len} and \code{sum} examples,
and the effect of \textbf{Recommendation (I)} to avoid mutable data on results
for \code{len} and \code{rev}.

\subsection{Loop Programs}
\label{sect:experiments:loops}
In our main experiment, we consider a larger set of common list-manipulating
tasks (such as checking if all/one element of a list is greater than a bound,
retrieving a list element by index, finding the index of a value, and computing
the maximum value).
Descriptions of all tasks are shown in \rSC{sect:examples} in the appendix.
We do not show results for the \asmName and \asmfxaName models, which always
fail.
We set the parameters for the remaining models to $\maxInt=32$ where possible
($\maxInt=\maxStack=34$ for the others), the length of the prefix
to 1, the length of the closure / loop body to 3 and the length of the suffix to
2.
Again, these parameters are slightly larger than required by the largest
program to be learned.

\begin{table}[H]
  \vspace{-1.5ex}
  \caption{Success ratios for full set of tasks.}
  \label{tab:loop-results}
  \vspace{-2ex}
  \hspace*{-1.75ex}
  \resizebox{.5\textwidth}{!}{
    \input{loopy_tbl.tex}

  }
  \vspace{-3ex}
\end{table}

The results for our experiments on these tasks are shown in
\rTab{tab:loop-results}.
Note the changed results of the examples from \rSC{sect:experiments:simple},
as the change in model parameters has increased the size of the program space
from $\sim 10^7$ to $\sim 10^{20}$.
The comparison to the \asmloopName model show the value of built-in
iteration and aggregation patterns.
The choice between immutable and mutable registers is less clear here,
seemingly dampened by other influences.
An inspection of the generated programs (eg. \rF{fig:last2-cmbmut} in the
appendix) reveals that mutability of registers can sometimes be exploited to
find elegant solutions.
Overall, it may be effective to combine both approaches, using a few mutable
``scratch value'' registers \emph{and} immutable default output registers for
each statement.

\subsection{Comparison with NRAM}
\label{sect:experiments:nram}
Our hypothesis was that the NRAM controller would fail to
generalize when trained on a small set of input-output examples.
As we believe that programming by example use cases usually operate on small
numbers of examples, we explicitly tested this hypothesis on the most simple
list-processing task len.
While not considered in \cite{Kurach15}, it is slightly simpler than
the \textbf{ListK} and \textbf{ListSearch} tasks that are classified as ``Hard
Tasks'' there.
We note that while very different, the NRAM model implements some of our
recommendations:
The RNN-like structure imposes a basic loop structure, and the output of each
module (i.e., instruction in our setting) is stored in a fixed register that is
immutable during a loop iteration.

\begin{figure}[t]
\begin{tikzpicture}
\definecolor{MinColor}{rgb}{0,0,1}
\definecolor{IntopsColor}{rgb}{0,1,0}
\definecolor{IntopsTwiceColor}{rgb}{1,0,0}

\pgfplotsset{
  min/.style={color=MinColor,mark=square*},
  intops/.style={color=IntopsColor,mark=square},
  intopsTwice/.style={color=IntopsTwiceColor,mark=*},
  every axis plot/.append style={line width=1pt},
}

\begin{axis}[
                 height=5cm,
                 width=.5\textwidth,
                 xlabel={Size of training set},
                 ylabel={Success ratio},
                 xmin=0, xmax=200,
                 ymin=0, ymax=50,
                 axis on top,
                 ytick={0,10,20,30,40,50},
                 yticklabels={0.0,,0.2,,0.4,,0.6,,0.8,,1.0},
                 xtick={10,20,30,40,50,60,70,80,90,100,110,120,130,140,150,160,170,180,190,200},
                 xticklabels={,,,,50,,,,,100,,,,,150,,,,,200},
                 enlargelimits=0.025,
                 legend entries={{min},{all},{all$^2$}},
                 legend cell align={left},
                ]
    \addplot [min] table {nram_len_min_setting4.data};
    \addplot [intops] table {nram_len_intops_setting91.data};
    \addplot [intopsTwice] table {nram_len_intopsTwice_setting91.data};
\end{axis}
\end{tikzpicture}
\vspace{-5ex}
\caption{Success ratio of NRAM on len}
\label{fig:nram-len-results}
\vspace{-4ex}
\end{figure}
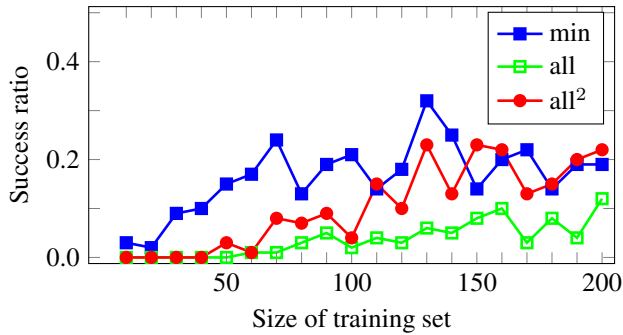

For our experiments, we simplified the NRAM model significantly and only
provided the modules \texttt{READ}, \texttt{ZERO}, \texttt{ONE}, \texttt{INC},
\texttt{ADD}, and \texttt{DEC} operating on integers.
Most notably, the absence of \texttt{WRITE} means that the heap remains
unchanged during program execution.
We considered three related models: 
 ``min'', in which \emph{only} the modules required in each iteration of a
 perfect solution are available,\footnote{For len, this was \texttt{INC},
 \texttt{INC}, \texttt{READ}.}
 ``all'' in which all modules are available once,\footnote{Note that this is
     the most challenging setting, as this requires the controller to choose
     different instructions in alternating iterations: One setting to advance
     the list pointer, and one to increment the length counter.} and
 ``all$^2$'', in which all modules are available twice.
We fixed the maximal length of input lists to 5, and the maximal integer to 11.

For each model, we performed an extensive random hyperparameter search
(choosing the optimizer, learning rate, momentum, size of the LSTM cell in the
NRAM controller, gradient noise, gradient clipping parameters, entropy bonus,
dropout probability), using 20 random restarts on 200 input-output pairs, with
validation and test sets of (disjoint) 50 examples each.
We stopped training after 200 epochs, or if the accuracy on the validation set
reached 100\% (most successful runs stopped after few epochs).
A run was counted as successful if the accuracy of the discretized model on
the test was 100\%, i.e., if the trained model successfully generalized to
unseen data of the same size.
For the best hyperparameter settings, we then varied the size of the training
set from $200$ down to $10$ in increments of $10$, keeping the size of the
validation set at a quarter of the size of the training set.
The results are displayed in \rF{fig:nram-len-results}.
We note that only the ``min'' model, where the NRAM controller chooses between
only
600 syntactically different programs in each iteration, has any success with
training sets smaller than 50.
Thus, our experiments confirm our hypothesis that NRAM-like models fail to
generalize from little data.


%% file: loopy_simple_tbl.tex
\newcolumntype{.}{D{.}{.}{3.2}}
\makeatletter
\newcolumntype{Z}[3]{>{\mathversion{nxbold}\DC@{#1}{#2}{#3}}c<{\DC@end}}
\makeatother
    \begin{tabular}{@{}l........@{}}
\toprule
\multicolumn{1}{l}{Program} & \multicolumn{1}{r}{\cmbName} & \multicolumn{1}{r}{\cmbmutName} & \multicolumn{1}{r}{\cmbuntName} & \multicolumn{1}{r}{\cmbuntmutName} & \multicolumn{1}{r}{\asmName} & \multicolumn{1}{r}{\asmfxaName} & \multicolumn{1}{r}{\asmloopName}  & \multicolumn{1}{r}{\lambdaSName} \\
\midrule
len & \multicolumn{1}{Z{.}{.}{3.2}}{100.00} & 75.00 & \multicolumn{1}{Z{.}{.}{3.2}}{100.00} & 43.67 & 0.00 & 0.00 & 15.67 & 100.00\\
rev & 48.33 & 32.67 & 46.33 & 41.33 & 0.00 & 0.00 & \multicolumn{1}{Z{.}{.}{3.2}}{86.33} & 100.00 \\
sum & \multicolumn{1}{Z{.}{.}{3.2}}{91.67} & 41.00 & 88.33 & 30.67 & 0.00 & 0.00 & 32.67 & 100.00 \\

\bottomrule
\end{tabular}

%% file: loopy_tbl.tex
\newcolumntype{.}{D{.}{.}{3.2}}
\makeatletter
\newcolumntype{Z}[3]{>{\mathversion{nxbold}\DC@{#1}{#2}{#3}}c<{\DC@end}}
\makeatother
    \begin{tabular}{@{}l......@{}}
\toprule
\multicolumn{1}{l}{Program} & \multicolumn{1}{r}{\cmbName} & \multicolumn{1}{r}{\cmbmutName} & \multicolumn{1}{r}{\cmbuntName} & \multicolumn{1}{r}{\cmbuntmutName} & \multicolumn{1}{r}{\asmloopName} & \multicolumn{1}{r}{\lambdaSName} \\
\midrule
$\text{len}$ & \multicolumn{1}{Z{.}{.}{3.2}}{98.67} & 96.33 & 0.67 & 0.33 & 0.00 & 100.00 \\
$\text{rev}$ & \multicolumn{1}{Z{.}{.}{3.2}}{18.00} & 10.33 & 2.67 & 8.33 & 9.67 & 100.00 \\
$\text{sum}$ & 38.00 & \multicolumn{1}{Z{.}{.}{3.2}}{38.33} & 1.00 & 0.00 & 10.00 & 100.00 \\
\midrule
$\text{allGtK}$ & 0.00 & 0.00 & 0.00 & \multicolumn{1}{Z{.}{.}{3.2}}{0.33} & 0.00 & 100.00 \\
$\text{exGtK}$ & \multicolumn{1}{Z{.}{.}{3.2}}{3.00} & 1.00 & 0.67 & 0.00 & 0.67 & 100.00\\
findLastIdx & \multicolumn{1}{Z{.}{.}{3.2}}{0.33} & 0.00 & 0.00 & 0.00 & 0.00 & 0.00\\
getIdx & \multicolumn{1}{Z{.}{.}{3.2}}{1.00} & 0.00 & 0.00 & 0.00 & 0.00 & 0.00\\
last2 & 0.00 & 8.00 & 0.00 & 2.00 & \multicolumn{1}{Z{.}{.}{3.2}}{23.00} & 0.00\\
$\text{mapAddK}$ & \multicolumn{1}{Z{.}{.}{3.2}}{100.00} & 98.00 & \multicolumn{1}{Z{.}{.}{3.2}}{100.00} & 95.67 & 0.00 & 100.00 \\
$\text{mapInc}$ & \multicolumn{1}{Z{.}{.}{3.2}}{99.67} & 98.00 & 99.33 & 97.00 & 0.00 & 100.00 \\
$\text{max}$ & 2.33 & \multicolumn{1}{Z{.}{.}{3.2}}{5.67} & 0.00 & 0.00 & 0.33 & 100.00\\
$\text{pairwiseSum}$ & 43.33 & 32.33 & \multicolumn{1}{Z{.}{.}{3.2}}{43.67} & 33.67 & 0.00 & 100.00\\
$\text{revMapInc}$ & 0.00 & 0.67 & 0.00 & 0.00 & \multicolumn{1}{Z{.}{.}{3.2}}{6.33} & 100.00\\

\bottomrule
\end{tabular}

%% file: discussion.tex
\section{Discussion and Future Work}

We have discussed a range of modeling choices for end-to-end differentiable
programming languages and made four design recommendations.
Empirically, we have shown these recommendations to significantly improve the
success ratio of learning programs from input/output examples, and we expect
these results to generalize to similar models attempting to learn programs.

In this paper, we only consider list-manipulating programs, but we are interested
in supporting more data structures, such as arrays (which should be a
straightforward extension) and associative maps.
We also only support loops over lists at this time, but are interested in
extending our models to also have built-in support for loops counting up to (and
down from) integer values.
A generalization of this concept would be an extension allowing the learning
and use of recursive functions.
Recursion is still more structured than raw goto calls, but more flexible than
the combinators that we currently employ.
An efficient implementation of recursion is a challenging research
problem, but it could allow significantly more complex programs to be learned.
Modeling recursion in an end-to-end differentiable language could allow us to build
libraries of (learned) differentiable functions that can be used in later
synthesis problems.

However, we note that with few exceptions on long straight-line code,
\lambdaSName performs better than all of our considered models, and is able to
synthesize programs in milliseconds.
We see the future of differentiable programming languages in areas in which
deterministic tools are known to perform poorly, such as the integration of
perceptual data, priors and ``soft'' side information such as natural language
hints about the desired functionality.


%% file: appendix.tex
\clearpage
\section{Appendix}

\subsection{Experiment Tasks}
\label{sect:examples}
  \begin{center}
  \begin{tabular}{@{}lp{11cm}@{}}
    \toprule
    Name & Description \\
    \midrule
    \code{len} & Return the length of a list. \\
    \code{rev} & Reverse a list. \\
    \code{sum} & Sum all elements of a list. \\
    \midrule
    \code{allGtK} & Check if all elements of a list are greater than $k$. \\
    \code{exGtK} & Check if at least one element of a list is greater $k$. \\
    \code{findLastIdx} & Find the index of the last list element which is equal to $v$. \\
    \code{getIdx} & Return the $k$th element of a list. \\
    \code{last2} & Return the 2nd to last element of a list. \\
    \code{mapAddK} & Compute list in which $k$ has been added to each element of
                     the input list. \\
    \code{mapInc} & Compute list in which each element of the input list has
                    been incremented. \\
    \code{max} & Return the maximum element of a list. \\
    \code{pairwiseSum} &  Compute list where each element is the sum of the corresponding elements
                          of two input lists.\\
    \code{revMapInc} & Reverse a list and increment each element. \\
    \bottomrule
  \end{tabular}
  \end{center}

Our example tasks for loop based programs. ``Simple'' tasks are above the line.

\subsection{Combinators}
\label{sect:appendix:combinators}

Semantics of \code{foldli}, \code{mapi}, \code{zipwithi} in a Python-like
language:

\begin{minipage}{0.5\textwidth}
\begin{algorithmic}
\Function{foldli}{$list, acc, func$}
\State $idx \gets 0$
\For{$ele$ \textbf{in} $list$}
\State $acc \gets func(acc, ele, idx)$
\State $idx \gets idx + 1$
\EndFor
\Return $acc$
\EndFunction
\end{algorithmic}
\end{minipage}%
\begin{minipage}{0.5\textwidth}
  \begin{algorithmic}
  \Function{mapi}{$list, func$}
  \State $idx \gets 0$
  \State $ret \gets [~]$
  \For{$ele$ \textbf{in} $list$}
  \State $ret \gets append(ret, func(ele, idx))$
  \State $idx \gets idx + 1$
  \EndFor
  \Return $ret$
  \EndFunction
\end{algorithmic}
\end{minipage}
\begin{algorithmic}
  \Function{zipwithi}{$list_1, list_2, func$}
  \State $idx \gets 0$
  \State $ret \gets [~]$
  \For{$ele_1, ele_2$ \textbf{in} $list_1, list_2$}
  \State $ret \gets append(ret, func(ele_1, ele_2, idx))$
  \State $idx \gets idx + 1$
  \EndFor
  \Return $ret$
  \EndFunction
\end{algorithmic}

\subsection{Selected Solutions}
\label{sect:appendix:exampleResults}

We show example results of our training in Figs.
\ref{fig:allGtK-cmbmutunt}-\ref{fig:sum-asmloop}.
Note that these are the actual results produced by our system, and have only
been slightly edited for typesetting.
Finally, we have colored statements that a simple program analysis can identify
as not contributing to the result in \textcolor{gray}{gray}.

\newcommand{\Let}[2]{\ensuremath{\textbf{let}\ {#1} = {#2}\ \textbf{in}}}
\newcommand{\Assign}[2]{\ensuremath{{#1} \gets {#2}}}
\newcommand{\Ite}[3]{\ensuremath{\textbf{if}\ {#1}\ \textbf{then}\ {#2}\ \textbf{else}\ {#3}}}
\newcommand{\StartFoldli}[3]{\State $\textbf{let}\ {#1} = \code{foldli}\ {#2}\ {#3}\ (\lambda\ ele\ acc\ idx \rightarrow$}
\newcommand{\StartFoldliM}[3]{\State \ensuremath{\Assign{#1}{\code{foldli}\ {#2}\ {#3}\ (\lambda\ ele\ acc\ idx \rightarrow}}}
\newcommand{\StartMapi}[2]{\State $\textbf{let}\ {#1} = \code{mapi}\ {#2}\ (\lambda\ ele\ idx \rightarrow$}
\newcommand{\StartZipWithi}[3]{\State $\textbf{let}\ {#1} = \code{zipWithi}\ {#2}\ {#3}\ (\lambda\ ele_1\ ele_2\ idx \rightarrow$}
\newcommand{\EndCmb}[1]{\ensuremath{{#1})\ \textbf{in}}}
\newcommand{\EndCmbM}[1]{\ensuremath{{#1})}}
\newcommand{\CmbIndent}[0]{\qquad}
%
\begin{figure*}[h]
\begin{algorithmic}
  \State $r_0 \gets l$
  \State $r_1 \gets k$
  \State \textcolor{gray}{$r_2 \gets r_0 \lor r_0$}
  \State $r_1 \gets \code{foldli}\ r_0\ r_0\ (\lambda\ ele\ acc\ idx \rightarrow$ 
  \State \CmbIndent $r_0 \gets ele > r_1$
  \State \CmbIndent \textcolor{gray}{$r_2 \gets \code{head}\ acc$}
  \State \CmbIndent $r_2 \gets r_0 \land acc$
  \State \CmbIndent $r_2)$
  \State $r_2 \gets r_1 \land r_0$
  \State \textcolor{gray}{$r_1 \gets r_1$}
  \State \Return $r_2$
\end{algorithmic}
\caption{A solution to \code{allGtK} in the C model. Code in gray is dead.}
\label{fig:allGtK-cmbmutunt}
\end{figure*}

\begin{figure*}
  \centering
  \begin{minipage}{0.5\textwidth}
    \centering
\begin{algorithmic}
\State \Let{r_0}{l}
\State \Let{r_1}{k}
\State \textcolor{gray}{\Let{r_2}{(r_0 = r_1)}}
\StartFoldli{r_3}{r_0}{r_0}
\State \CmbIndent \Let{c_0}{acc \lor acc}
\State \CmbIndent \Let{c_1}{ele > r_1}
\State \CmbIndent \Let{c_2}{c_0 \lor c_1}
\State \CmbIndent \EndCmb{c_2}
\State \textcolor{gray}{\Let{r_4}{r_3 \lor r_3}}
\State \textcolor{gray}{\Let{r_5}{r_3 \land r_2}}
\State \Return $r_4$
\end{algorithmic}
  \end{minipage}%
  \begin{minipage}{0.5\textwidth}
    \centering
\begin{algorithmic}
  \State $r_0 \gets l$
  \State $r_1 \gets k$
  \State $r_2 \gets r_2 = r_1$
  \For{$ele$ \textbf{in} $r_0$}
  \State \textcolor{gray}{$r_0 \gets \Ite{r_2}{ele}{r_1}$}
  \State $r_0 \gets ele > r_1$
  \State $r_2 \gets r_2 \lor r_0$
  \EndFor
  \State $r_2 \gets r_2 \lor r_0$
  \State \textcolor{gray}{$r_1 \gets r_2 \lor r_2$}
  \State \Return $r_2$
\end{algorithmic}
  \end{minipage}
  \caption{Solutions to \code{exGtK} in the C+T+I and A+L models.}
\end{figure*}

\begin{figure*}
\begin{algorithmic}
\State \Let{r_0}{l}
\State \Let{r_1}{e}
\State \Let{r_2}{r_0 + 1}
\StartFoldli{r_3}{r_0}{r_2}
\State \CmbIndent \textcolor{gray}{\Let{c_0}{\Ite{r_2}{idx}{r_1}}}
\State \CmbIndent \Let{c_1}{(r_1 = ele)}
\State \CmbIndent \Let{c_2}{\Ite{c_1}{idx}{acc}}
\State \CmbIndent \EndCmb{c_2}
\State \textcolor{gray}{\Let{r_4}{r_3 + 1}}
\State \textcolor{gray}{\Let{r_5}{r_2}}
\State \Return $r_3$
\end{algorithmic}
\caption{A solution to \code{findLastIdx} in the C+T+I model.}
\end{figure*}

\begin{figure*}
\begin{algorithmic}
\State \Let{r_0}{l}
\State \Let{r_1}{k}
\State \Let{r_2}{\code{head}\  r_0}
\StartFoldli{r_3}{r_0}{r_2}
\State \CmbIndent \Let{c_0}{(r_1 = idx)}
\State \CmbIndent \Let{c_1}{\Ite{c_0}{ele}{acc}}
\State \CmbIndent \textcolor{gray}{\Let{c_2}{\Ite{idx}{idx}{c_0}}}
\State \CmbIndent \EndCmb{c_1}
\State \textcolor{gray}{\Let{r_4}{\code{head}\  r_0}}
\State \textcolor{gray}{\Let{r_5}{\code{tail}\ r_0}}
\State \Return $r_3$
\end{algorithmic}
\caption{A solution to \code{getIdx} in the C+T+I model.}
\end{figure*}

\begin{figure*}
  \centering
  \begin{minipage}{0.5\textwidth}
    \centering
    \begin{algorithmic}
      \State \Assign{r_0}{l}
      \State \Assign{r_1}{0}
      \State \textcolor{gray}{\Assign{r_2}{nil}}
      \StartFoldliM{r_2}{r_0}{r_1}
      \State \CmbIndent \textcolor{gray}{\Assign{r_0}{acc}}
      \State \CmbIndent \Assign{r_2}{r_1}
      \State \CmbIndent \Assign{r_1}{ele}
      \State \CmbIndent \EndCmb{r_2}
      \State \textcolor{gray}{\Assign{r_0}{r_2 + r_2}}
      \State \textcolor{gray}{\Assign{r_0}{r_0 + 1}}
      \State \Return $r_2$
    \end{algorithmic}
  \end{minipage}%
  \begin{minipage}{0.5\textwidth}
    \centering 
    \begin{algorithmic}
      \State \Assign{r_0}{l}
      \State \Assign{r_1}{0}
      \State \Assign{r_2}{\code{cons}\ r_0\ r_0}
      \For{$(ele_1, ele_2)$ \textbf{in} $(r_0, r_2)$}
      \State \Assign{r_2}{\Ite{r_2}{ele_2}{ele_2}}
      \State \textcolor{gray}{\Assign{r_1}{r_2 - 1}}
      \State \textcolor{gray}{\Assign{r_1}{\code{head}\  r_0}}
      \EndFor
      \State \textcolor{gray}{\Assign{r_1}{\Ite{r_1}{r_2}{r_0}}}
      \State \textcolor{gray}{\Assign{r_0}{\Ite{r_2}{r_0}{r_1}}}
      \State \Return $r_2$
    \end{algorithmic}
  \end{minipage}
  \caption{Solutions to \code{last2} in the C+T and A+L models.}
  \label{fig:last2-cmbmut}
  \label{fig:last2-asmloop}
\end{figure*}

 
\begin{figure*}
\begin{algorithmic}
\State \Let{r_0}{l}
\State \textcolor{gray}{\Let{r_1}{0}}
\State \textcolor{gray}{\Let{r_2}{\code{tail}\ r_0}}
\StartFoldli{r_3}{r_0}{r_0}
\State \CmbIndent \Let{c_0}{idx + 1}
\State \CmbIndent \textcolor{gray}{\Let{c_1}{\Ite{r_1}{r_2}{c_0}}}
\State \CmbIndent \textcolor{gray}{\Let{c_2}{c_0 = ele}}
\State \CmbIndent \EndCmb{c_0}
\State \textcolor{gray}{\Let{r_4}{\Ite{r_2}{r_3}{r_3}}}
\State \textcolor{gray}{\Let{r_5}{r_3 + r_2}}
\State \Return $r_3$
\end{algorithmic}
\caption{A solution to \code{len} in the C+T+I model.}
\end{figure*}

\begin{figure*}
\begin{algorithmic}
\State \Let{r_0}{l}
\State \Let{r_1}{k}
\State \textcolor{gray}{\Let{r_2}{\Ite{r_1}{r_0}{r_0}}}
\StartMapi{r_3}{r_0}
\State \CmbIndent \textcolor{gray}{\Let{c_0}{ele - 1}}
\State \CmbIndent \textcolor{gray}{\Let{c_1}{c_0 - 1}}
\State \CmbIndent \Let{c_2}{r_1 + ele}
\State \CmbIndent \EndCmb{c_2}
\State \textcolor{gray}{\Let{r_4}{ r_3}}
\State \textcolor{gray}{\Let{r_5}{ r_3}}
\State \Return $r_3$
\end{algorithmic}
\caption{A solution to \code{mapAddK} in the C+T+I model.}
\end{figure*} 

\begin{figure*}
\begin{algorithmic}
\State \Let{r_0}{l}
\State \textcolor{gray}{\Let{r_1}{0}}
\State \textcolor{gray}{\Let{r_2}{r_0}}
\StartMapi{r_3}{r_0}
\State \CmbIndent \textcolor{gray}{\Let{c_0}{\Ite{r_1}{ele}{acc}}}
\State \CmbIndent \Let{c_1}{ele + 1}
\State \CmbIndent \textcolor{gray}{\Let{c_2}{ r_1}}
\State \CmbIndent \EndCmb{c_1}
\State \textcolor{gray}{\Let{r_4}{\code{cons}\  r_3\ r_0}}
\State \textcolor{gray}{\Let{r_5}{ r_4}}
\State \Return $r_3$
\end{algorithmic}
\caption{A solution to \code{mapInc} in the C+T+I model.}
\end{figure*}

\begin{figure*}
  \centering
  \begin{minipage}{0.5\textwidth}
    \centering
    \begin{algorithmic}
      \State \Let{r_0}{l}
      \State \textcolor{gray}{\Let{r_1}{0}}
      \State \textcolor{gray}{\Let{r_2}{\code{tail}\ r_0}}
      \StartFoldli{r_3}{r_0}{r_0}
      \State \CmbIndent \Let{c_0}{acc > ele}
      \State \textcolor{gray}{\CmbIndent \Let{c_1}{ acc}}
      \State \CmbIndent \Let{c_2}{\Ite{c_0}{acc}{ele}}
      \State \CmbIndent \EndCmb{c_2}
      \State \textcolor{gray}{\Let{r_4}{r_2 - 1}}
      \State \textcolor{gray}{\Let{r_5}{\code{head}\  r_2}}
      \State \Return $r_3$
    \end{algorithmic}
  \end{minipage}%
  \begin{minipage}{0.5\textwidth}
    \centering
    \begin{algorithmic}
      \State \Assign{r_0}{l}
      \State \Assign{r_1}{0}
      \State \Assign{r_2}{r_0 - 1}
      \For{$(ele_1, ele_2)$ \textbf{in} $(r_0, r_0)$}
      \State \textcolor{gray}{\Assign{r_0}{ ele_1}}
      \State \Assign{r_1}{ele_1 > r_2}
      \State \Assign{r_2}{\Ite{r_1}{ele_1}{r_2}}
      \EndFor
      \State \textcolor{gray}{\Assign{r_0}{r_0 + r_0}}
      \State \textcolor{gray}{\Assign{r_0}{\code{cons}\  r_2\ r_2}}
      \State \Return $r_2$
    \end{algorithmic}
  \end{minipage} 
  \caption{Solutions to \code{max} in the C+T+I and A+L models.}
\end{figure*}
 
\begin{figure*}
\begin{algorithmic}
\State \Let{r_0}{l_1}
\State \Let{r_1}{l_2}
\State \textcolor{gray}{\Let{r_2}{\Ite{r_1}{r_0}{r_1}}}
\StartZipWithi{r_3}{r_1}{r_0}
\State \CmbIndent \Let{c_0}{ele1 + ele2}
\State \CmbIndent \textcolor{gray}{\Let{c_1}{ele2 - 1}}
\State \CmbIndent \textcolor{gray}{\Let{c_2}{idx - 1}}
\State \CmbIndent \EndCmb{c_0}
\State \textcolor{gray}{\Let{r_4}{\Ite{r_0}{r_3}{r_1}}}
\State \textcolor{gray}{\Let{r_5}{\Ite{r_4}{r_2}{r_1}}}
\State \Return $r_3$
\end{algorithmic}
\caption{A solution to \code{pairwiseSum} in the C+T+I model.}
\end{figure*}

\begin{figure*}
  \centering
  \begin{minipage}{0.5\textwidth}
    \centering
    \begin{algorithmic}
      \State \Let{r_0}{l}
      \State \Let{r_1}{0}
      \State \textcolor{gray}{\Let{r_2}{\code{cons}\  r_0\ r_0}}
      \StartFoldli{r_3}{r_0}{r_1}
      \State \CmbIndent \textcolor{gray}{\Let{c_0}{\code{cons}\  ele\ acc}}
      \State \CmbIndent \textcolor{gray}{\Let{c_1}{\code{cons}\  acc\ acc}}
      \State \CmbIndent \Let{c_2}{\code{cons}\  ele\ acc}
      \State \CmbIndent \EndCmb{c_2}
      \State \textcolor{gray}{\Let{r_4}{\Ite{r_2}{r_3}{r_2}}}
      \State \textcolor{gray}{\Let{r_5}{\code{cons}\  r_4\ r_3}}
      \State \Return $r_3$
    \end{algorithmic}
  \end{minipage}%
  \begin{minipage}{0.5\textwidth}
    \centering
    \begin{algorithmic}
      \State \Assign{r_0}{l}
      \State \Assign{r_1}{0}
      \State \Assign{r_2}{\code{tail}\ r_1}
      \For{$ele_1$ \textbf{in} $r_0$}
      \State \textcolor{gray}{\Assign{r_1}{\code{cons}\  ele_2\ r_0}}
      \State \Assign{r_2}{\code{cons}\  ele_1\ r_2}
      \State \textcolor{gray}{\Assign{r_1}{\code{tail}\ r_0}}
      \EndFor
      \State \textcolor{gray}{\Assign{r_0}{\code{tail}\ r_0}}
      \State \textcolor{gray}{\Assign{r_0}{\code{cons}\  r_2\ r_1}}
      \State \Return $r_2$
    \end{algorithmic}
  \end{minipage}
  \caption{Solutions to \code{rev} in the C+T+I and A+L models.}
\end{figure*}

 
\begin{figure*}
  \centering
  \begin{minipage}{0.5\textwidth}
    \centering
    \begin{algorithmic}
      \State \Assign{r_0}{l}
      \State \textcolor{gray}{\Assign{r_1}{0}}
      \State \textcolor{gray}{\Assign{r_1}{\code{tail}\ r_0}}
      \StartFoldliM{r_2}{r_0}{r_2}
      \State \CmbIndent \Assign{r_2}{ele + 1}
      \State \textcolor{gray}{\CmbIndent \Assign{r_0}{\code{cons}\  r_2\ acc}}
      \State \CmbIndent \Assign{r_2}{\code{cons}\  r_2\ acc}
      \State \CmbIndent \EndCmbM{r_2}
      \State \textcolor{gray}{\Assign{r_1}{\code{cons}\  r_2\ r_1}}
      \State \textcolor{gray}{\Assign{r_0}{\code{cons}\  r_2\ r_0}}
      \State \Return $r_2$
    \end{algorithmic}
  \end{minipage}%
  \begin{minipage}{0.5\textwidth}
    \centering
    \begin{algorithmic}
      \State \Assign{r_0}{l}
      \State \textcolor{gray}{\Assign{r_1}{0}}
      \State \textcolor{gray}{\Assign{r_1}{1}}
      \For{$ele_1$ \textbf{in} $r_0$}
      \State \Assign{r_0}{ele1 + 1}
      \State \textcolor{gray}{\Assign{r_1}{1}}
      \State \Assign{r_2}{\code{cons}\  r_0\ r_2}
      \EndFor
      \State \textcolor{gray}{\Assign{r_1}{\code{cons}\  r_0\ r_2}}
      \State \textcolor{gray}{\Assign{r_1}{\code{cons}\  r_0\ r_2}}
      \State \Return $r_2$
    \end{algorithmic}
  \end{minipage}
  \caption{Solutions to \code{revMapInc} in the C+T and A+L models.}
\end{figure*}

\begin{figure*}
  \centering
  \begin{minipage}{0.5\textwidth}
    \centering
\begin{algorithmic}
\State \Let{r_0}{l}
\State \textcolor{gray}{\Let{r_1}{0}}
\State \textcolor{gray}{\Let{r_2}{r_0}}
\StartFoldli{r_3}{r_0}{r_0}
\State \CmbIndent \textcolor{gray}{\Let{c_0}{acc + r_0}}
\State \CmbIndent \Let{c_1}{acc + ele}
\State \CmbIndent \textcolor{gray}{\Let{c_2}{\Ite{r_0}{idx}{r_1}}}
\State \CmbIndent \EndCmb{c_1}
\State \textcolor{gray}{\Let{r_4}{r_2 + 1}}
\State \textcolor{gray}{\Let{r_5}{r_3 - 1}}
\State \Return $r_3$
\end{algorithmic}
\end{minipage}%
  \begin{minipage}{0.5\textwidth}
    \centering
\begin{algorithmic}
\State \Assign{r_0}{l}
\State \textcolor{gray}{\Assign{r_1}{0}}
\State \textcolor{gray}{\Assign{r_1}{\Ite{r_2}{r_1}{r_0}}}
\For{$ele_1$ \textbf{in} $r_0$}
  \State \Assign{r_2}{ele_1 + r_2}
  \State \textcolor{gray}{\Assign{r_1}{\code{cons}\  r_2\ r_0}}
  \State \textcolor{gray}{\Assign{r_1}{ele_1 + ele_2}}
\EndFor
\State \textcolor{gray}{\Assign{r_0}{r_1 + r_1}}
\State \textcolor{gray}{\Assign{r_0}{r_2 + 1}}
\State \Return $r_2$
\end{algorithmic}
\end{minipage}
\caption{Solutions to \code{sum} in the C+T+I and A+L models.}
\label{fig:sum-asmloop}
\end{figure*}